\documentclass[11pt]{article}

\usepackage[utf8]{inputenc}
\usepackage[margin=1in]{geometry}
\usepackage{graphicx}
\usepackage{amsmath}
\usepackage{amssymb}
\usepackage{rotating}
\usepackage{graphicx}
\usepackage{wrapfig}
\usepackage{lscape}
\usepackage{rotating}
\usepackage{epstopdf}
\usepackage{float}
\usepackage[nottoc]{tocbibind}
\usepackage{caption}
\usepackage{subcaption}
\usepackage{mathtools}
\usepackage[affil-it]{authblk}
\usepackage{siunitx}
\usepackage[colorlinks=true, allcolors=blue]{hyperref}
\usepackage{comment}

\begin{document}

\title{Frustrated total internal reflection of ultrasonic waves at a fluid-coupled elastic plate}

\author{André Lello de Almeida$^{a}$, Ming Huang$^a$, Peiyao Huang$^b$, Frederic Cegla$^a$, Bo Lan$^{a}$\footnote{\raggedright{\noindent Corresponding author: bo.lan@imperial.ac.uk.}}}
\affil{
$^a$ Department of Mechanical Engineering, Imperial College London, London SW7 2AZ, UK\\
$^b$ Department of Engineering, University of Cambridge, Trumpington Street, Cambridge CB2 1PZ, UK
}
\date{\vspace{-5ex}}
\maketitle

\begin{abstract}
A complete treatment regarding frustrated total internal reflection (FTIR) of ultrasonic waves is presented and validated against experiments, providing a theoretical explanation for the physics behind this phenomenon. Two different approaches are used to develop a theoretical model capable of studying transmission in fluid-coupled elastic plates. One is the multiple reflections approach (analogous to the study of FTIR in electromagnetic/optical waves), which is shown to have limited applicability for incident angles beyond the first critical angle. This prompted us to address the problem using the second, potentials-based approach, which is established and validated against experimental data with correct predictions for a thin air-coupled steel sheet. A relation between the transmitted power and the dispersion curves for guided waves in the plate is established, highlighting the two fundamental causes of FTIR in such systems. First, a thin plate, when compared to the wavelength of the wave incident on it, will always be subjected to the effects of FTIR. This is because the evanescent wave created inside the plate will never assume negligible values, thus always allowing transmission to the other side. Second, and more surprisingly, the excitation of the fundamental antisymmetric mode $A_0$ of the plate is shown to be a direct enabler of FTIR, even for thicker plates.
\end{abstract}

\section{Introduction}
\label{sec,intro}
The phenomenon of frustrated total internal reflection (FTIR) has been known to occur for electromagnetic waves for quite some time \cite{Zhu et al_1986}. It relates to the effect of light penetrating a medium with lower index of refraction and having the possibility of being transmitted through that same medium, whenever total internal reflection (TIR) is to be expected instead. Traditionally, in optics this happens if the penetrated medium is sufficiently thin, compared to the wavelength of incident light.

Many authors have studied this phenomenon and have tried to explain it within the evolving framework of optics. Newton \cite{Newton_1952} was the first to discover and experimentally investigate it, followed by Fresnel \cite{Fresnel_1823} and others. Although many of these scientists and physicists have explored the penetration of light into the second medium, only after some time the first reliable quantitative work was presented by Quincke \cite{Quincke_1866}. Later, a very complete treatment of FTIR was published in a paper by Hall \cite{Hall_1902}, based on a theoretical exposition by Drude \cite{Drude_1959}.

Hall \cite{Hall_1902} based all of his experimental and theoretical investigations on a three-medium system, where this third extra medium, with a low index of refraction, was used as a separation between the other two. This would allow him to measure the amount of light transmitted into the second medium, and consequently how this penetration was affected by the properties of the system. However, as stated in his paper, Hall could not find a way to experimentally detect this phenomenon when only two media were present.

Thus, further work to establish what we now call evanescent waves was pursued, culminating in the works of Eichenwald \cite{Eichenwald_1911}, Foersterling \cite{Foersterling_1937}, and Arzelies \cite{Arzelies_1946}. Another important contribution was that of the discovery of the Goos-H\"{a}nchen shift \cite{Goos and Hanchen_1947}, characterised by a lateral displacement of the reflected wave. The reasons behind this effect were discussed by Renard \cite{Renard_1964}, and presented very clearly in his theoretical treatment of the subject.

These advances and a much more comprehensive understanding of FTIR allowed the evolution of important applications in optics, such as in the field of spectroscopy and in the design of laser resonators \cite{Zhu et al_1986}. It seems obvious that this phenomenon should not manifest itself exclusively for electromagnetic waves, but also for elastic ones, such is the interdisciplinary nature of wave mechanics.

Besides the intellectual and scientific values, a better grasp of the underlying physics of the FTIR phenomenon for elastic waves will also be relevant when dealing with the practical aspects of ultrasound propagating through thin sheets, allowing for better non-destructive evaluation (NDE) of various engineering structures, like metal sheet composites or electrodes of Li-ion batteries. Therefore, we aspire to study this same effect when ultrasonic waves interact with boundaries, hoping that applications using these kinds of systems will be founded upon a better fundamental understanding of the behaviours of the reflected and transmitted waves. In addition to this and to the best of the knowledge of the authors, there is currently no literature regarding the study of FTIR in ultrasounds.

While FTIR only occurs for angles greater than the critical ones, this paper will analyse the whole problem of oblique incidence for any angle, using a complex wavenumber formulation to account for the possibility of modelling evanescent waves. Besides that, it is stipulated that FTIR only occurs when the thickness of the plate is sufficiently small. However, such a transition should be apparent in a theory that makes no prior considerations regarding the ratio of the thickness of the plate and the wavelength of the wave.

For the sake of simplicity, we will focus on the case of a solid isotropic plate submerged in a fluid. This will not only simplify the generation and reception of ultrasounds, since only longitudinal waves can travel in the fluid, but it will also guarantee (at least for the usual engineering materials) that the fluid is separated by a medium with a lower index of refraction, without any loss of generality.

We will start by discussing a first attempt at describing FTIR using multiple reflections, directly transplanting the formulation used in optics \cite{Zhu et al_1986}. However, as it will soon be shown, such an approach is rather impractical for the case of elastic waves, due to complications that arise when accounting for mode conversion between longitudinal and shear waves. The poor and computationally inefficient results produced by its numerical implementation led to the adoption of another theoretical formulation based on the use of potentials. The results obtained by this model will be analysed next, with the physical mechanisms behind FTIR being explained. Following this, experimental evidence is used to validate the theoretical formulation.

\section{Theoretical Framework}
\label{sec,theoretical_framework}

\subsection{Multiple reflections approach}
\label{sec,multiple_reflections}
A multiple reflections approach is first employed to get a completely analogous theory of elastic waves to that for electromagnetic ones. This might be the natural approach to consider given the close proximity of elastic and optical waves. With this in mind, some references are used to provide a contextual framework in which our own theoretical development could be built upon. The concise review of the subject provided by Zhu et al \cite{Zhu et al_1986}, the study on phase shifts accompanying the evanescent wave by Azzam \cite{Azzam_2006}, and the theoretical summary and experimental discussion in V\"{o}r\"{o}s and Johnsen's work \cite{Voros and Johnsen_2008} are all useful in the development of this model.

Starting with the Fresnel relations \cite{Born and Wolf_1980}, representing how the amplitude of the different waves change with the angle of incidence, it will be necessary to find equivalent relations for the amplitudes of elastic waves interacting with a boundary. Of course extensive literature already exists on this particular topic \cite{Graff_1975,Rose_1999}. After this, we will consider a ray-based approach to calculate the contribution of all possible paths the waves could take when reflecting inside the plate.

As an example, consider the path where the incident wave transmits into the solid as a longitudinal wave, reflecting twice as itself, and at the third and last reflection converts into a shear wave before coming out of the same side of the plate. The contribution to the final reflection coefficient of this path can be written as,

\begin{equation}
\label{eqn,path_contribution}
A_R = A_Il_{12}rl_{21}^2ml_{21}ls_{21}e^{3i\delta_L}e^{i\delta_S}
\end{equation}

\noindent where $A_I$ is the amplitude of the incident wave, $l_{12}$ represents the conversion to a longitudinal wave in the solid, $rl_{21}$ represents reflection of longitudinal waves inside the solid, $ml_{21}$ represents mode conversion from a longitudinal to a shear wave inside the solid, $ls_{21}$ represents the conversion from a shear wave in the solid to a wave in the fluid, $\delta_L$ is the phase shift of a longitudinal wave travelling between plate walls, and $\delta_S$ is the same phase shift, but for shear waves.

In theory, the infinite sum of all possible different contributions will converge and the final result will allow us to calculate the total transmission and reflection coefficients. However, mainly due to mode conversion between longitudinal and shear waves inside the plate, the problem turns out to be quite difficult, if possible at all, to solve analytically, using currently available mathematical tools. Nonetheless, we still believe that this method would eventually lead to the correct result, if the series is simplified into an analytical expression. To demonstrate this, equation (\ref{eqn,partial_wave_mode_conversion}) shows the final analytical result considering the existence of mode conversion,

\begin{equation}
\label{eqn,partial_wave_mode_conversion}
\frac{A_R}{A_I} = r_{12} + \frac{l_{12}ll_{21}}{\sqrt{1+M_l}} \frac{\left(1+\sqrt{1+M_l}\right)^2rl_{21}e^{2i\delta_L}}{4-\left(1+\sqrt{1+M_l}\right)^2rl_{21}^2e^{2i\delta_L}} + \frac{s_{12}ls_{21}}{\sqrt{1+M_s}}
\frac{\left(1+\sqrt{1+M_s}\right)^2rs_{21}e^{2i\delta_S}}{4-\left(1+\sqrt{1+M_s}\right)^2rs_{21}^2e^{2i\delta_S}} + M
\end{equation}

\noindent with $r_{12}$ representing the reflection at the boundary with the solid, $ll_{21}$ being the counterpart of $ls_{21}$ for longitudinal waves, $s_{12}$ and $rs_{21}$ being the counterparts of $l_{12}$ and $rl_{21}$ for shear waves, and $M_l$, $M_s$, and $M$ being factors that depend on mode conversion coefficients. If it is assumed that no mode conversion can occur, $M_l=M_s=M=0$, we get,

\begin{equation}
\label{eqn,partial_wave_no_conversion}
\frac{A_R}{A_I} = r_{12} + l_{12}ll_{21}\frac{rl_{21}e^{2i\delta_L}}{1-rl_{21}^2e^{2i\delta_L}} + s_{12}ls_{21}\frac{rs_{21}e^{2i\delta_S}}{1-rs_{21}^2e^{2i\delta_S}}
\end{equation}

\noindent which is very similar to analogous results obtained in optics \cite{Zhu et al_1986}. However, because $M$ has not a readily available analytical form, a finite series numerical solution is the only way we can possibly obtain results, though they happen to have limited utility in addressing our goal of discussing and understanding FTIR. Further discussion of the mathematical model used is provided in Appendix \ref{append,multiple_reflections}, its main equations have been omitted for the sake of brevity.

In the plots shown in Figure \ref{fig,partial_d}, results for two different incidence angles are plotted using the multiple reflections approach. For this purpose, we used a system consisting of an infinitely long copper plate coupled with water on both sides. The relevant properties of both materials are shown in Table \ref{tab,properties_copper_water}. The results obtained are in accordance to what was expected and give some insight into why those peaks manifest in those particular conditions.

\begin{table}[!t]
    \centering
    \caption{Density and elastic wave velocities for water and copper.}
    \begin{tabular}{c c c}
        \hline
        & Water (1) & Copper (2) \\ \hline
        Density $\rho$ (kg/m\textsuperscript{3}) & 1000 & 8900 \\
        Longitudinal Wave Velocity $c_L$ (m/s) & 1480 & 4660 \\
        Shear Wave Velocity $c_S$ (m/s) & - & 2260 \\
        \hline
    \end{tabular}
    \label{tab,properties_copper_water}
\end{table}

\begin{figure}[!b]
    \centering
    \includegraphics[width=1\linewidth]{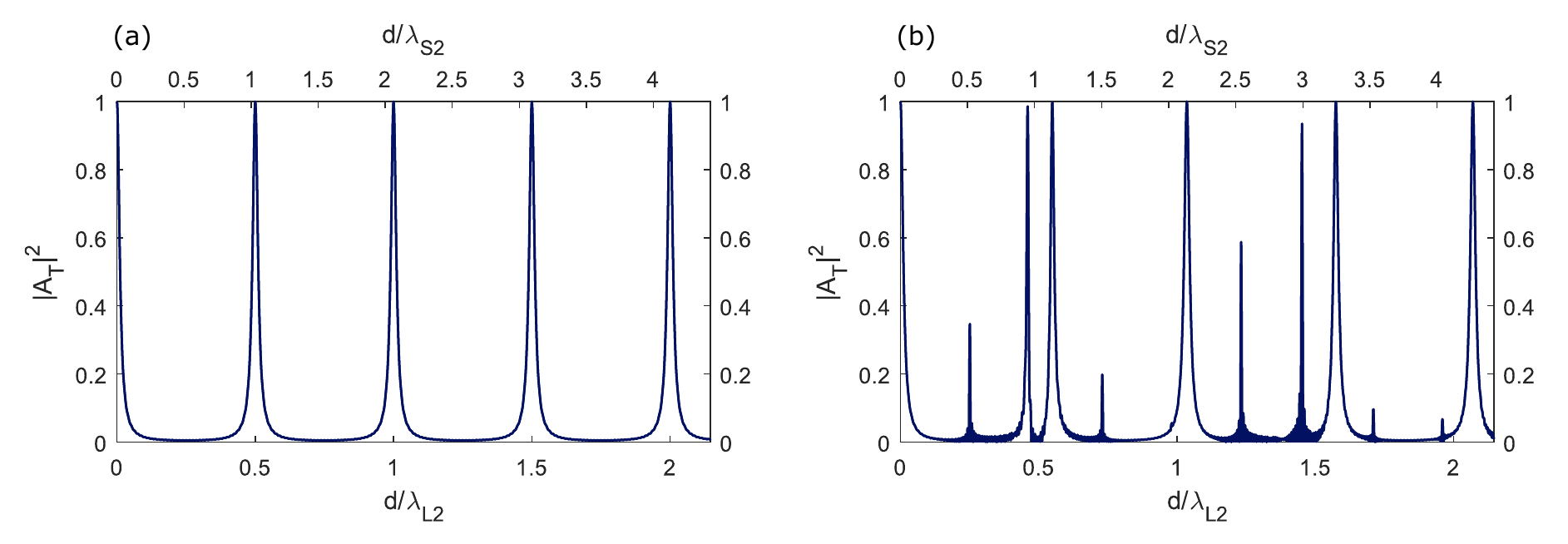}
    \caption{Transmitted power as a function of plate thickness at normal incidence (a) and for a fixed incidence angle of $5^\circ$ (b), using the multiple reflections approach. The horizontal bottom axis shows the ratio between plate thickness and the wavelength of longitudinal waves in the solid, whilst the top one represents the ratio between plate thickness and the wavelength of shear waves.}
    \label{fig,partial_d}
\end{figure}

First, in Figure \ref{fig,partial_d}(a) the results for normal incidence are shown. It is apparent that total transmission occurs when the thickness of the plate is an integer multiple of half the wavelength of longitudinal waves. This is a direct consequence of the constructive interference between multiply-reflected longitudinal waves bouncing between the two surfaces of the plate. Since no shear waves are excited in this case, no resonances should occur for integer half-wavelengths of shear waves, which can also be observed.

On the other hand, after a careful investigation of Figure \ref{fig,partial_d}(b), which shows results for angled incidence, it seems that peaks in transmission occur very close to values of thickness corresponding to half-integer multiples of both longitudinal and shear wavelengths inside the plate. All of this is further evidence that the coefficient of transmission is connected to the reverberation of waves inside the plate. The lateral displacement that affects the peaks is due to the oblique incidence in itself, as the waves travel slightly different distances than when they are normal to the boundaries.

Finally, whilst Figure \ref{fig,partial_d}(a) represents an exact solution, as all the shear and mode conversion terms vanish, the oblique incidence case can only be approximated using a finite sum. It is even possible to note this difference by the appearance of a small oscillating behaviour in Figure \ref{fig,partial_d}(b), whenever the transmitted power approaches zero. These oscillations can be understood as simple artefacts of using a finite sum to represent the analytically correct infinite one. In fact, increasing the number of terms used in the expansion does lead to a decrease in the amplitude of the oscillations.

The shortcomings of the multiple reflections approach are more clearly apparent for larger angles of incidence. Up to the first critical angle, $18.5^\circ$, the model behaves well, if we disregard the slight oscillations appearing near some transmission peaks. However, after this angle the multiple reflections approach completely breaks down, the transmitted power approaching $-\infty$ or $\infty$ for certain ranges of incidence angles. Thus, we can conclude that, due to a lack of an analytical solution and an unstable numerical implementation, the model only applies for a large plate thickness (when compared to the wavelength of waves inside the plate) and for angles of incidence before the first critical angle.

A possible reason why the multiple reflections approach did not work past the first critical angle, as we hoped, is related to the very nature of evanescent waves. In this ray-based approach, we simply consider the corresponding wavenumber as a complex number, whenever an evanescent wave is present. However, using complex wavenumbered rays may not give a suitable representation for these waves, not because it is wrong to do so, but because it does not specify how a complex wavenumbered wave would physically interact with a boundary. To explain it in other words, according to this model, these complex evanescent waves would travel and interact with the boundaries along complex directions in a real three-dimensional space, lacking a physical meaning.

These paragraphs highlight the difficulties in addressing FTIR using the multiple reflections approach, and demonstrate that it is imperative to have a theoretical framework capable of making predictions in the regimen where the thickness is very small compared to the wavelengths of the elastic waves present in the system. Besides, we are very much interested in studying all possible angles of incidence, especially after the first and second critical angles, where TIR is supposed to occur. Therefore, the logical next step is to consider an alternative approach to this problem.

\subsection{Wave field solution using potentials}
\label{sec,wave_field}
Instead of using rays to represent the propagation of waves in a mechanical system, it is possible to consider the full wave field by solving the three-dimensional equations of elasticity. This way, the evanescent waves will appear as a natural extension to the theory, taking away the need to understand \textit{a priori} how they reverberate inside the plate. Following derivations appearing in the works of Graff \cite{Graff_1975} and Rose \cite{Rose_1999}, we may represent the displacement field as potentials, by Helmholtz decomposition.

\begin{figure}[!t]
    \centering
    \includegraphics[scale=0.6]{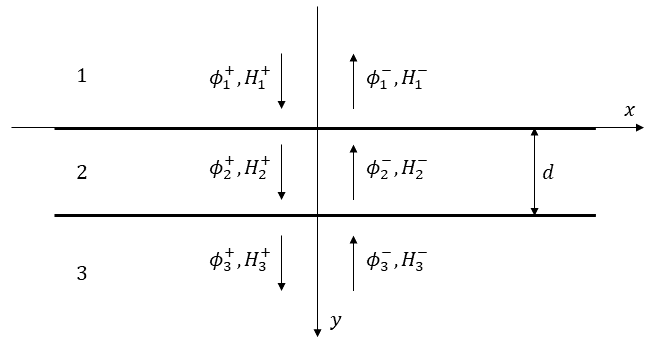}
    \caption{Diagram of a plate submerged in a fluid and the corresponding potentials for each region. Region 1 and 3 represent an inviscid fluid surrounding an isotropic solid plate, shown as region 2. Longitudinal waves are incoming only from region 1 into the upper boundary of region 2.}
    \label{fig,potentials_diagram}
\end{figure}

First, it is necessary to separate space into three different regions, with three different potential fields as shown in Figure \ref{fig,potentials_diagram}. We will have the incident region of the fluid (1), the inside of the solid plate (2), and the fluid on the other side (3). Therefore, the scalar and vector potentials, associated respectively with longitudinal and shear waves, can be written as travelling waves to the right \cite{Graff_1975},

\begin{equation}
\label{eqn,scalar_potential}
\phi_i=A_i^+e^{i(\xi x+\alpha_i y-\omega t)}+A_i^-e^{i(\xi x-\alpha_i y-\omega t)}
\end{equation}

\begin{equation}
\label{eqn,vector_potential}
H_2=B_2^+e^{i(\xi x+\beta_2 y-\omega t)}+B_2^-e^{i(\xi x-\beta_2 y-\omega t)}
\end{equation}

\noindent where $i$ can go from 1 to 3, $A_1^+=1$ if we assume a unit incident wave, $A_1^-=R$ representing the reflection coefficient, $A_3^+=T$ representing the transmission coefficient, and $A_3^-=0$ as there is no incident wave from the other side. $H_2$ represents the out-of-plane component of the vector potential and the other two components vanish; since shear waves cannot propagate inside the fluid, $H_1$ and $H_3$ are both zero. Besides, $\xi$ represents the x-component of the wavenumber, equal for every wave, by Snell's law. $\alpha_i$ and $\beta_2$ are, respectively, the y-components of the wavenumbers of the longitudinal and shear waves. $\omega$ is the angular frequency.

We assume that the fluid media 1 and 3 are the same and have the longitudinal wavenumber of $k_{L1}=\omega/c_{L1}$. The solid medium 2 in the middle have the longitudinal and shear wavenumber of $k_{L2}=\omega/c_{L2}$ and $k_{S2}=\omega/c_{S2}$. In this case, the aforementioned wavenumber components are given by,

\begin{equation}
\label{eqn,x_wavenumber}
\xi=k_{L1}\sin\theta
\end{equation}

\begin{equation}
\label{eqn,y_wavenumber}
\alpha_{1,3}=k_{L1}\cos\theta,\quad\quad \alpha_2=\sqrt{k_{L2}^2-\xi^2},\quad\quad \beta_2=\sqrt{k_{S2}^2-\xi^2}
\end{equation}

\noindent where $k_i$ represents the wavenumber of each wave and $\theta$ is the angle of incidence.

To solve for the remaining unknown parameters $A_i^+$, $A_i^-$, $B_2^+$, and $B_2^-$, it is imperative that we apply the correct boundary conditions (BC) to this problem. After a careful analysis of the system, continuity of normal displacement and stress should be accomplished, as well as the vanishing of shear stresses at both plate boundaries, since no shear stress can be sustained inside an inviscid fluid.

So, from the definition of potentials used, it is known that the displacement and stress fields can be derived explicitly as \cite{Graff_1975},

\begin{equation}
\label{eqn,x_displacement}
u_x=\frac{\partial\phi}{\partial x}+\frac{\partial H}{\partial y}
\end{equation}

\begin{equation}
\label{eqn,y_displacement}
u_y=\frac{\partial\phi}{\partial y}-\frac{\partial H}{\partial x}
\end{equation}

\begin{equation}
\label{eqn,normal_stress}
\tau_{yy}=\left(\lambda+2\mu\right)\left(\frac{\partial^2\phi}{\partial x^2}+\frac{\partial^2\phi}{\partial y^2}\right)-2\mu\left(\frac{\partial^2\phi}{\partial x^2}+\frac{\partial^2H}{\partial x\partial y}\right)
\end{equation}

\begin{equation}
\label{eqn,shear_stress}
\tau_{xy}=\mu\left(2\frac{\partial^2\phi}{\partial x\partial y}+\frac{\partial^2H}{\partial y^2}-\frac{\partial^2H}{\partial x^2}\right)
\end{equation}

\noindent where $\lambda$ and $\mu$ are the Lam\'e parameters of the corresponding material, and they are related to the longitudinal and shear wave speeds by $c_L=\sqrt{(\lambda+2\mu)/\rho}$ and $c_S=\sqrt{\mu/\rho}$; for a fluid material, $\mu=0$ and its longitudinal wave speed is $c_L=\sqrt{\lambda/\rho}$.

Using equations (\ref{eqn,x_displacement}) to (\ref{eqn,shear_stress}), it is possible to enforce the six BC defined previously,

\begin{equation}
\label{eqn,bc_y_0}
u_{y1}=u_{y2}\quad,\quad \tau_{yy1}=\tau_{yy2}\quad,\quad \tau_{xy2}=0\quad\quad:\quad\quad y=0
\end{equation}

\begin{equation}
\label{eqn,bc_y_d}
u_{y2}=u_{y3}\quad,\quad \tau_{yy2}=\tau_{yy3}\quad,\quad \tau_{xy2}=0\quad\quad:\quad\quad y=d
\end{equation}

\noindent which can be solved for the missing coefficients. It is interesting to see how the phase shift applied in the multiple reflections approach appears here naturally, due to half of the BC being defined at $y=d$, where $d$ is the thickness of the plate.

The system of six equations can be written explicitly if we substitute the expressions for the displacement and stress fields in the BC defined in equations (\ref{eqn,bc_y_0}) and (\ref{eqn,bc_y_d}). The six equations are written in matrix form as,

\begin{equation}
\label{eqn,matrix_equation_1}
    \begin{bmatrix}
    \alpha_1 & \alpha_2 & -\alpha_2 & -\xi & -\xi & 0 \\
    \rho_1\omega^2 & -C_1 & -C_1 & C_3 & -C_3 & 0 \\
    0 & C_2 & -C_2 & C_4 & C_4 & 0 \\
    0 & -\alpha_2e^{i\alpha_2d} & \alpha_2e^{-i\alpha_2d} & \xi e^{i\beta_2d} & \xi e^{-i\beta_2d} & \alpha_1e^{i\alpha_1d} \\
    0 & C_1e^{i\alpha_2d} & C_1e^{-i\alpha_2d} & -C_3e^{i\beta_2d} & C_3e^{-i\beta_2d} & -\rho_1\omega^2e^{i\alpha_1d} \\
    0 & C_2e^{i\alpha_2d} & -C_2e^{-i\alpha_2d} & C_4e^{i\beta_2d} & C_4e^{-i\beta_2d} & 0
    \end{bmatrix}
    \begin{bmatrix} R \\ A_2^+ \\ A_2^- \\ B_2^+ \\ B_2^- \\ T \end{bmatrix} =
    \begin{bmatrix} \alpha_1 \\ -\rho_1\omega^2 \\ 0 \\ 0 \\ 0 \\ 0 \end{bmatrix}
\end{equation}

\noindent with,

\begin{equation}
\label{eqn,matrix_equation_2}
    C_1=(\rho_2\omega^2-2\mu_2\xi^2),\; C_2=2\mu_2\xi\alpha_2, \; C_3=2\mu_2\xi\beta_2, \; C_4=\mu_2(k_{S2}^2-2\xi^2)
\end{equation}

For a given angle of incidence $\theta$, solving equation (\ref{eqn,matrix_equation_1}) leads to the solutions for the six unknown coefficients. Therefore, by computing $T$ for every value of $\theta$, it is possible to establish the behaviour of the transmission coefficient as a function of the angle of incidence.

\section{Transmission Coefficient Analysis}
\label{sec,transmission_analysis}

\subsection{Transmitted power as a function of incidence angle and plate thickness}
\label{sec,transmitted_power}
In Figure \ref{fig,full_d}, four plots of the transmission coefficient using the potentials solution are shown. To obtain these results, the same system of a cooper plate coupled with water (properties given in Table \ref{tab,properties_copper_water}) was considered.

\begin{figure}[!t]
    \centering
    \includegraphics[width=1\linewidth]{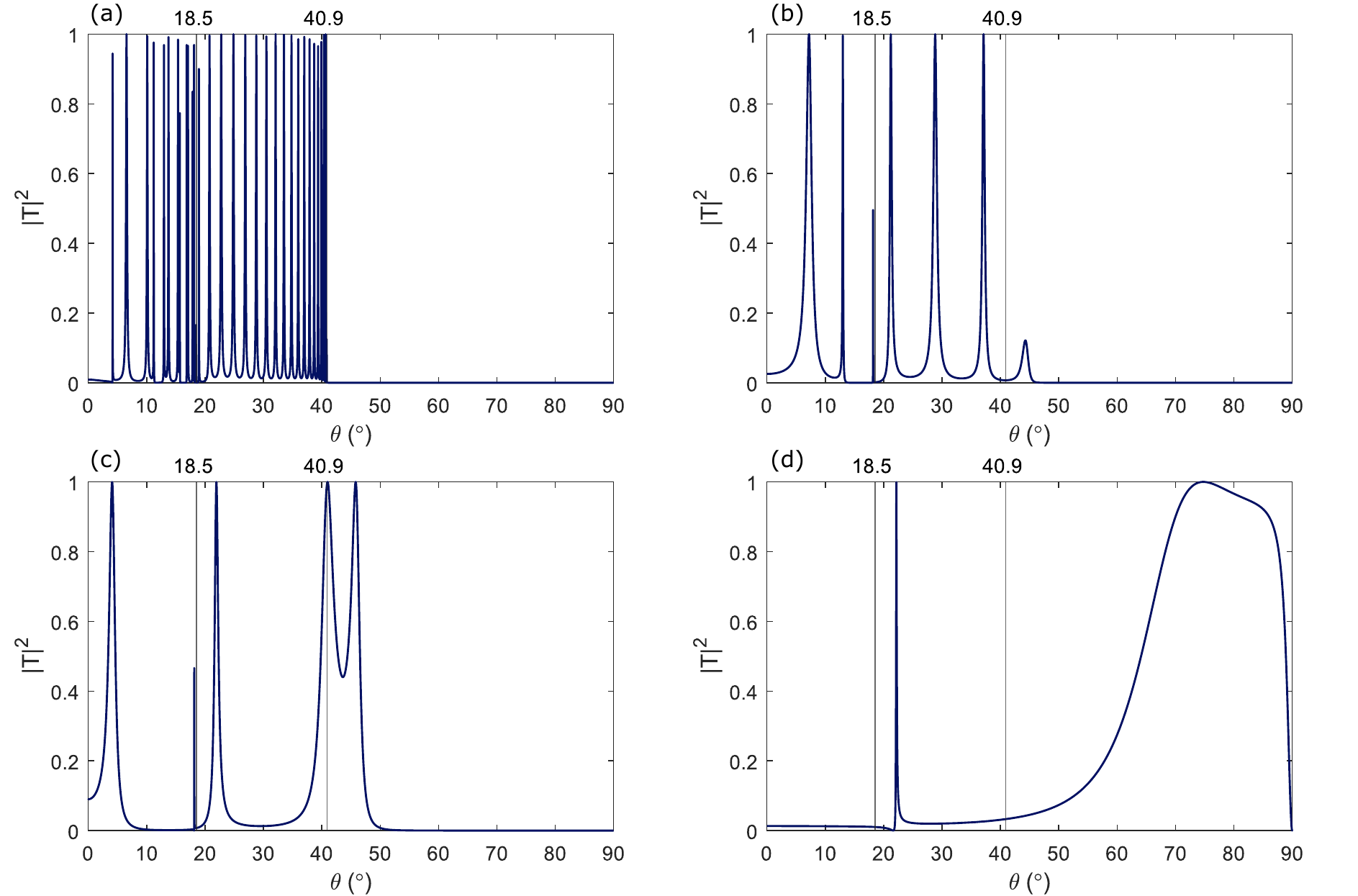}
    \caption{Transmitted power as a function of angle of incidence using the potential wave field solution for $d = 5$ mm (a), $d = 1$ mm (b), $d = 0.5$ mm (c) , and $d = 0.1$ mm (d). Plot (a) represents a case where $d \gg \lambda_{L2}$, whilst (b) stands for situations where $d \approx \lambda_{L2}$. Plot (d) stands for an example where $d \ll \lambda_{L2}$, since $\lambda_{L2} = 0.932$ mm for an incident frequency of $5$ MHz. The two vertical lines represent the first, $18.5^\circ$, and second, $40.9^\circ$, critical angles.}
    \label{fig,full_d}
\end{figure}

First of all, it is important to note that the transmitted power varies between 0 and 1, as in our model there can be no energy stored in the plate in a steady-state condition. Besides, summing the reflected and transmitted powers gives a value of exactly 1 for all possible angles, further satisfying conservation of energy.

It is possible to observe a tendency on the behaviour of the transmitted power after the second critical angle ($40.9^\circ$) as the thickness of the plate gets smaller. Figure \ref{fig,full_d}(a) represents a case where the thickness is substantially larger than the wavelength of waves inside the plate and no waves can be transmitted through the plate after the shear critical angle. It seems that TIR dominates the system when the plate thickness is large compared to the wavelength of waves travelling inside that same plate.

However, looking at Figure \ref{fig,full_d}(b), where the thickness is roughly the same as the wavelength of longitudinal waves inside the solid, a slight peak in transmission exists right after the second critical angle. This peak increases both in magnitude and width for a smaller and smaller plate thickness, as we can see in the potentials-based solution graphs of Figures \ref{fig,full_d}(c) and \ref{fig,full_d}(d). The existence of all those other transmission peaks is of course an interesting part of the behaviour of the system, but to truly grasp the reasons behind the occurrence of FITR, it will be crucial to better understand the physical origin of the particular peak appearing after the second critical angle.

With this in mind, let us turn our attention to the specific scenario shown in Figure \ref{fig,full_d}(b), so that we may isolate that peak and study that particular solution to the equations defined in the previous section. The best way to visualise how the waves are behaving is to plot the displacement field for that particular angle of incidence and thickness. Therefore, the real part of this field is represented in Figure \ref{fig,displacement_scholte}. It is important to note that the displacement was turned dimensionless to account for the fact that we are using a unit incident potential wave ($U_y = \frac{u_y}{k_{L1}\cos\theta}$, $U_x = \frac{u_x}{k_{L1}\sin\theta}$).

From Figure \ref{fig,displacement_scholte}(a), it is possible to see that the continuity condition on the normal displacement is indeed satisfied, and that the exponential behaviour of the evanescent waves is apparent inside the metal. Of course, continuity of transverse displacement was not a necessary condition to be satisfied and that is why the surfaces are not continuous in Figure \ref{fig,displacement_scholte}(b). To the right of the plate ($y > 1$ mm), the transmitted wave is clearly visible and travelling at an angle. To the left ($y < 0$ mm), the displacement field is more complicated as it results from the interference between the incident and reflected waves.

\begin{figure}[!t]
    \centering
    \includegraphics[width=1\linewidth]{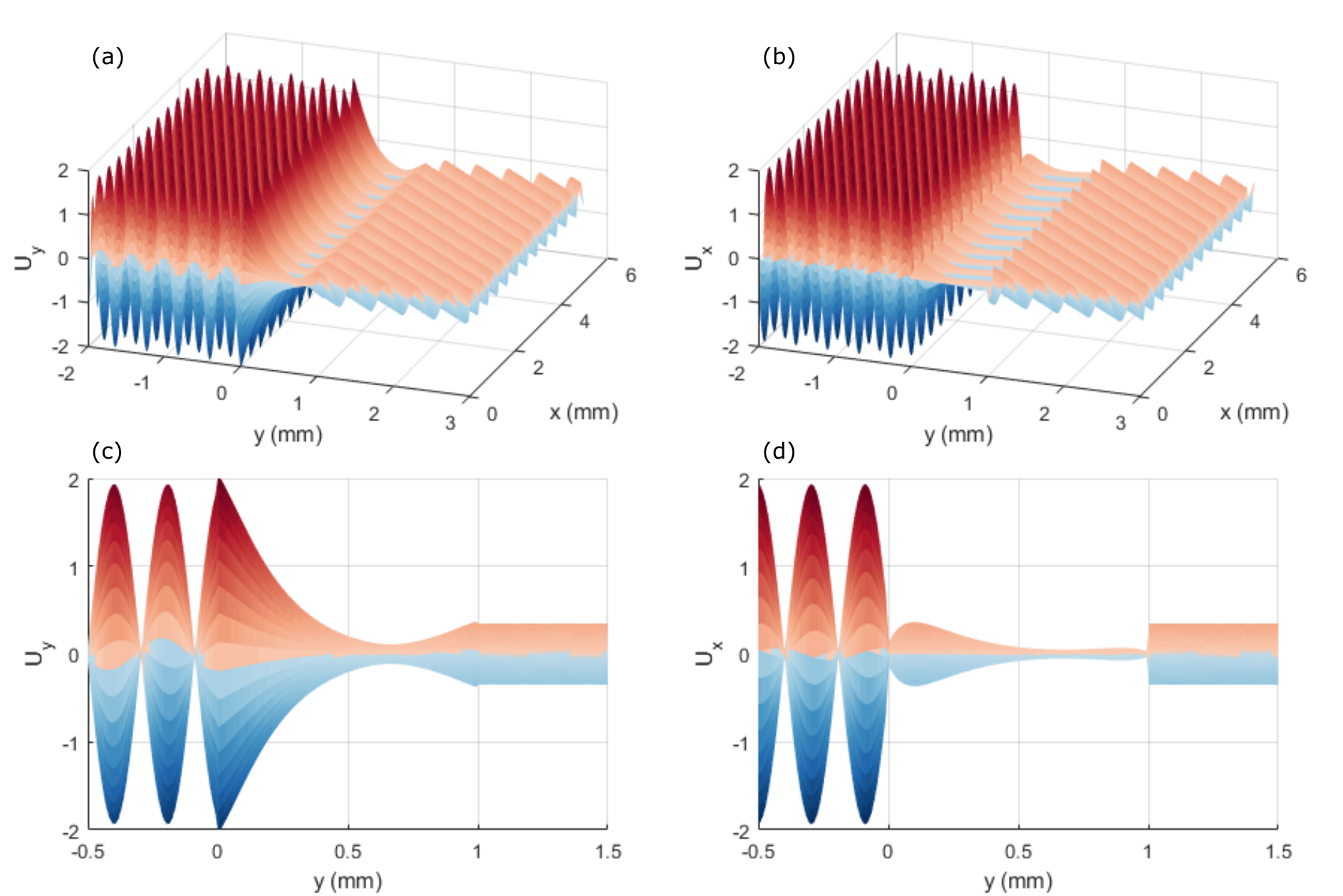}
    \caption{Normalised displacement field for $d=1$ mm and $\theta=44.3^\circ$. Plots (a) and (c) show the normalised normal displacement field, while (b) and (d) show the normalised transverse displacement field. Besides this, plots (c) and (d) show a detailed profile view of the interaction with and within the plate. The incoming wave is moving towards positive $x$ and $y$ and the plate is located between $y = 0$ mm and $y = 1$ mm.}
    \label{fig,displacement_scholte}
\end{figure}

Now, it is suspected that transmission peaks in this system are caused by some kind of reverberations inside the plate. This means that in order to understand the formation of this specific peak, we might be interested in understanding exactly how the field behaves inside the solid. So, Figures \ref{fig,displacement_scholte}(c) and \ref{fig,displacement_scholte}(d) show a profile close-up view of the displacement field in the plate.

The resulting displacement field is quite unique and provides a lot of clues as to what is really happening in the plate. In Figure \ref{fig,displacement_scholte}(c), the normal displacement varies as a decreasing exponential, passing by a minimum before increasing exponentially until continuity has been achieved on both surfaces. As for the transverse displacement in Figure \ref{fig,displacement_scholte}(d), it increases rapidly from zero before reaching a maximum and decaying slowly. Besides this, it is a bit difficult to notice, but both displacements seem to be out-of-phase.

All these characteristics are remarkably consistent with the displacement field caused by a Scholte wave \cite{Stoneley_1924,Scholte_1942}, which may not be so surprising after all. Indeed, one can derive the dispersion equation for a Scholte wave by solving the determinant of a matrix very similar to our own, but assuming that the wavenumbers in both potentials are complex, instead of real numbers \cite{Rose_1999}.

Of course, because this is not a half-space problem, since there is an interaction with fluids on both sides of the plate, the particular peak we are studying really represent the interaction between Scholte waves on both surfaces of the plate. This is the reason why we can see an exponentially decreasing behaviour followed by an exponentially increasing one in Figure \ref{fig,displacement_scholte}(c).

Another way of thinking about this interaction borrows from the fundamental theory of Lamb guided waves on a free plate \cite{Lamb_1917}. The lowest mode of propagation of guided waves corresponds exactly to an interaction between Rayleigh waves \cite{Rayleigh_1885} on both surfaces of the plate, which sounds very similar to what is happening in our case, exchanging these Rayleigh waves by Scholte waves.

Therefore, making the following leap by analogy, one might understand that the transmission peak is caused by the excitation of the lowest mode of propagation of guided waves on the plate. In other words, we might describe it as the fundamental antisymmetric mode of guided waves in a fluid-coupled elastic plate, $A_0$ \cite{Graff_1975}. Or using another nomenclature, the quasi-Scholte mode of that particular plate \cite{Cegla et al_2005}.

After this discussion, one must acknowledge that the other peaks might also be connected to the propagation of guided waves in the plate. In fact, in the following sections we will provide some proof corroborating this position.

\subsection{Relation to guided waves in a fluid-coupled elastic plate}
\label{sec,guided_waves}
It might be interesting to analyse the transmitted power dependence with the whole spectrum of possible values of both plate thickness and incidence angle. This would mean representing the transmitted power as a function of these two variables, which is shown in Figure \ref{fig,dispersion_curves}.

\begin{figure}[!b]
    \centering
    \includegraphics[scale=0.9]{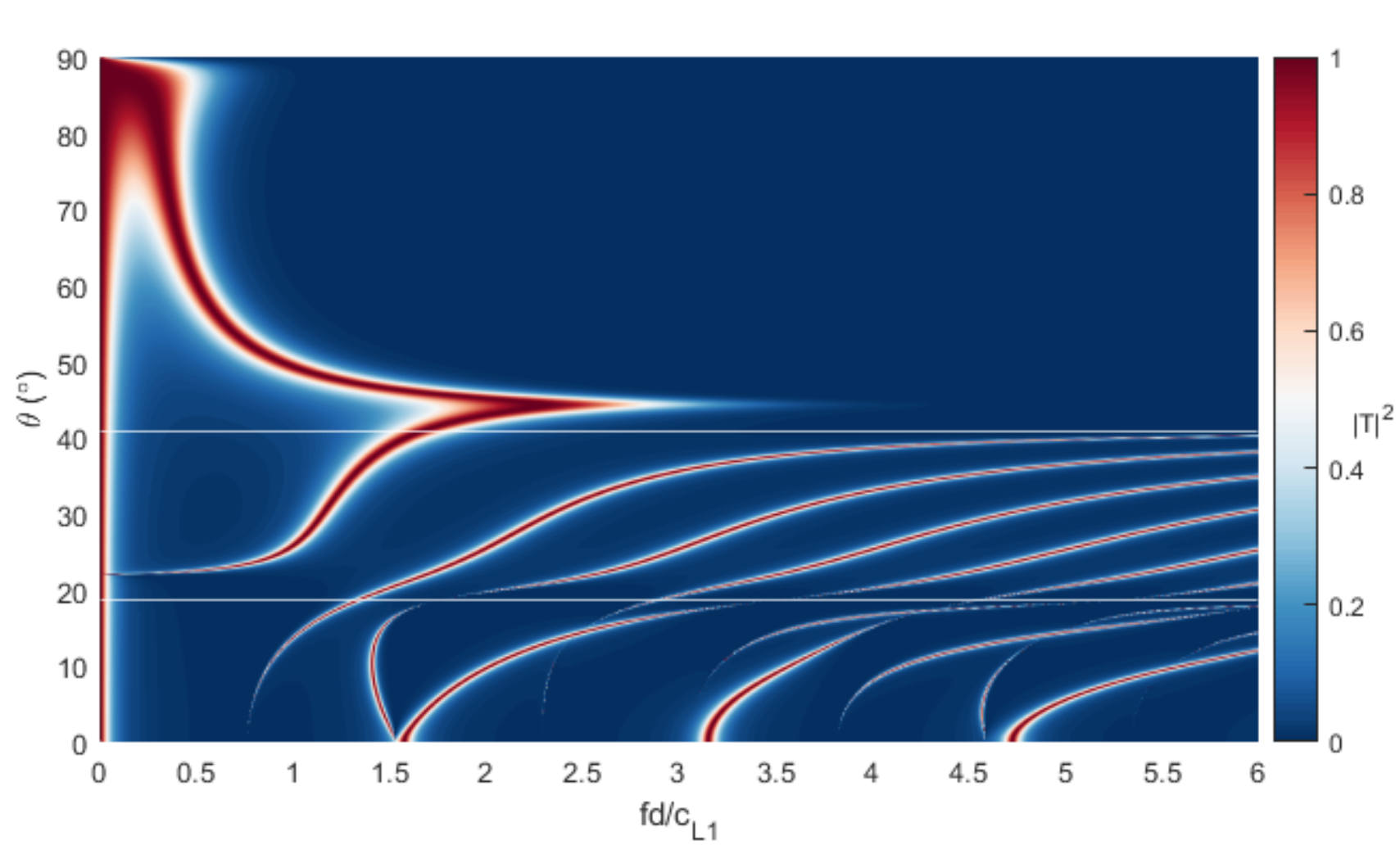}
    \caption{Transmitted power as a function of plate thickness and incidence angle for a plate submerged in a fluid. The two horizontal white lines represent the first, $18.5^\circ$, and second, $40.9^\circ$, critical angles.}
    \label{fig,dispersion_curves}
\end{figure}

In this plot, the y-axis is the angle of incidence and the x-axis is a dimensionless number representing the ratio between the plate thickness and the wavelength of the incident wave. White horizontal lines show the critical angles of the system we have been considering so far: a copper plate submerged in water.

These curves look very similar to dispersion curves at first sight, but using angle of incidence as the y-axis variable instead of guided wave velocity. It is possible in fact to compare these curves to the ones shown in relevant literature \cite{Chimenti and Rokhlin_1990}, studying these altered Lamb type guided waves in a fluid-coupled elastic plate. Performing the transformation shown in equation (\ref{eqn,velocity_angle_incidence}) between these two variables, we can change the domain of the plot from the velocity-frequency domain to the angle of incidence-frequency domain, as desired \cite{Rose_1999},

\begin{equation}
\label{eqn,velocity_angle_incidence}
\theta = \arcsin\left(\frac{c_{L1}}{c}\right)
\end{equation}

\noindent where $c_{L1}$ is the velocity of elastic waves in the fluid and $c$ is the phase velocity of guided waves in the plate.

Compared to plots of dispersion curves in the angle of incidence-frequency domain found in the literature \cite{Rose_1999}, our lines of high transmitted power resemble these guided wave modes very closely. The implications of this are very exciting and corroborates previous work \cite{Chimenti and Rokhlin_1990} in this area, which established that zeroes in the reflection coefficient corresponded to the excitation of guided waves in the plate and that the presence of a fluid would shift the position of these zeroes. Besides that, Cremer \cite{Cremer_1942} had already suggested that the impedance of a fluid-coupled plate would effectively vanish if the trace velocity of the incident wave was equal to the velocity of travelling waves inside the plate.

For a particular plate thickness, and angle of incidence and frequency of an incoming wave, guided waves may be excited or not in the plate. When they are, the wave effectively gets absorbed, being radiated to the other side due to the existence of a coupling between the fluid and the plate. So, we can conclude that transmission peaks occur as a phenomenon of impedance matching, where the plate becomes effectively transparent to incoming waves, if they arrive in such a way that guided waves are excited in the solid.

Analysing the plot of Figure \ref{fig,dispersion_curves} more closely, it is possible to divide it into three different regions, where their respective defining characteristics agree with our own intuition of what should happen in each a case. First, for a very low  thickness-wavelength ratio, the transmitted power is completely independent of the angle of incidence and close to 1, effectively asserting the well-known observation that for large wavelengths, the wave essentially ignores such a thin plate. Also, in the limit when the thickness goes to 0, the plate is non existing, so that the entirety of the incoming wave should transmit, which can be readily seen in our plot. This vertical red stripe clearly indicates a region where FTIR can occur at any angle of incidence.

In the second region, close to a thickness ratio of 1, there are at least two peaks, one corresponding to the fundamental antisymmetric mode $A_0$, or fundamental flexural mode of the plate, and another representing the fundamental symmetric mode $S_0$, or fundamental extensional mode of the plate. These two modes are also the only ones that can exist after the second critical angle, with the flexural one appearing always after this angle. This indicates that these two modes, but mainly $A_0$, are the ones responsible for the occurrence of FTIR in thicker plates.

At last, for very high frequencies or very thick plates, the peak after the second critical angle fades away, leaving only a high number of peaks before this point. This number of peaks also increases with increasing thickness-wavelength ratio, as the number of excitable modes available becomes larger. The fading of the $A_0$ or $S_0$ peak, also signifies the regimen where TIR occurs as expected. As shown in the work of Chimenti and Rokhlin \cite{Chimenti and Rokhlin_1990} this fading away is a direct result of exchanges among branches of the complex dispersion spectrum. As the density of the fluid gets smaller it approaches an effective vacuum, reaching, in the limit, a state where the plate vibrates freely, exactly the condition necessary for regular Lamb waves to propagate. In this situation, both fundamental modes fade only at much higher frequencies, implying that for dense solids, when compared to the density of the coupling fluid, FTIR might occur even for a rather large plate thickness.

Of course, the knowledge that FTIR is associated with the fundamental flexural mode of a plate gives some enticing possibilities with regards to inversion problems and other applications. It may be possible in the future to develop a methodology, such that by measuring FTIR transmission peaks, the properties of the fluid or the plate can be deduced. We could, for example, use a simpler representation of $A_0$, such as with a Mindlin plate theory \cite{Mindlin_1951}, to establish a relation between the transmission peaks and the properties of the media involved, without the complications of the full dispersion equation.

Furthermore, the fact that using potentials is a very well known method of achieving solutions in elasticity problems \cite{Graff_1975,Rose_1999} and that our equations are similar to other references studying fluid-coupled elastic plates \cite{Chimenti and Rokhlin_1990}, allow an optimistic point of view of the model used here to predict the transmission peaks, and consequently FTIR.

\subsection{Phase behaviour of the transmitted waves}
\label{sec,transmitted_phase}
\begin{figure}[!b]
    \centering
    \includegraphics[width=1\linewidth]{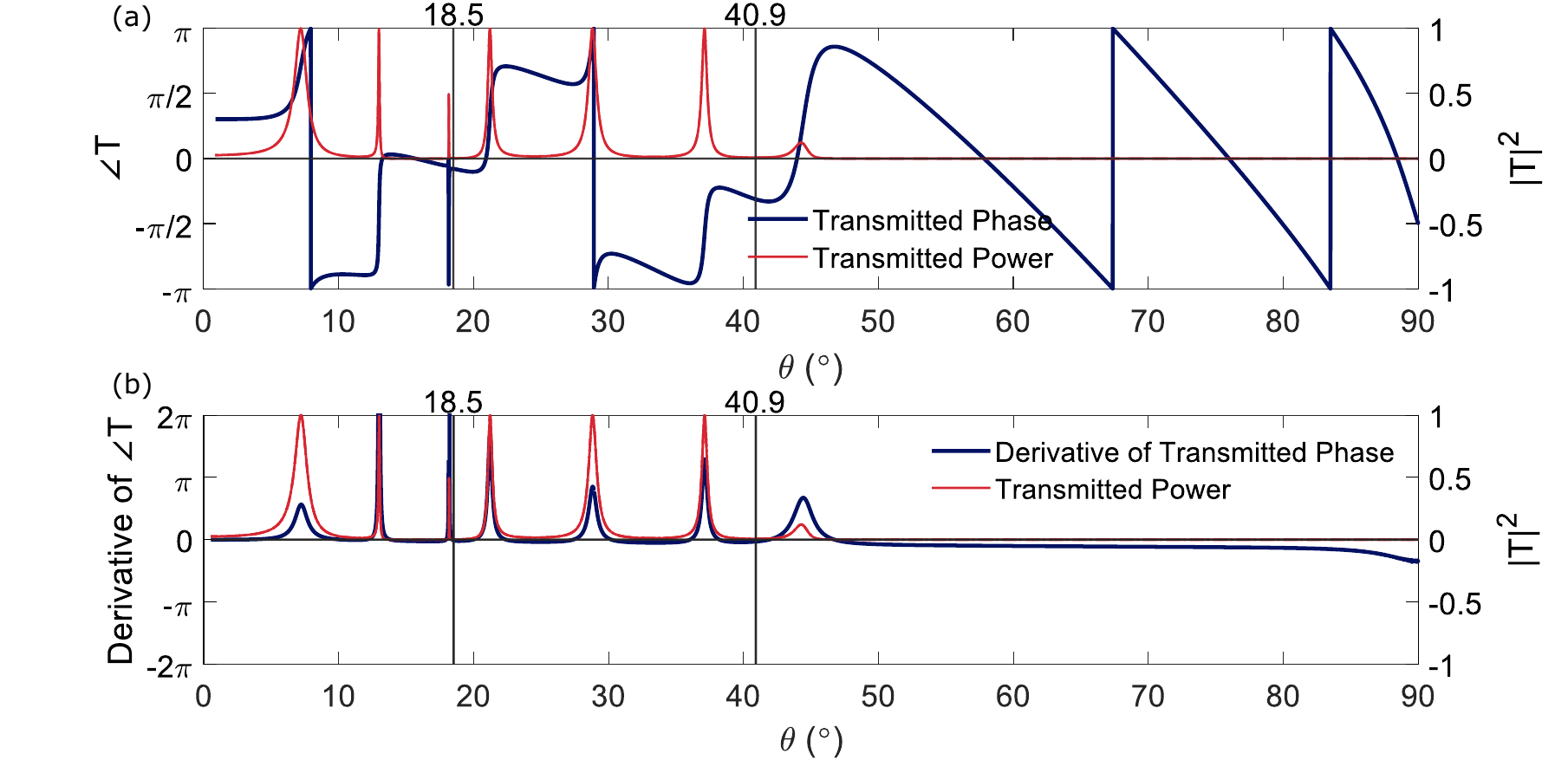}
    \caption{Transmitted power and phase as functions of angle of incidence for $d = 1$ mm (a). The derivative of the transmitted phase with respect to the incidence angle is also shown against the transmitted power (b). The two vertical lines represent the first, $18.5^\circ$, and second, $40.9^\circ$, critical angles.}
    \label{fig,phase_d}
\end{figure}

All of our theoretical analysis up until now discussed exclusively the behaviour of the transmission coefficient, indirectly by considering the transmitted power, without any regards for how the phase changes. This new information might give further insights into the excitation of guided waves in a fluid-coupled plate. Therefore, transmitted power and phase for a particular plate thickness are shown in Figure \ref{fig,phase_d}(a).

From the plot it is easily seen that whenever there is a peak in transmitted power, the phase varies very rapidly, followed by a slight decrease in its rate of change. This rapid variation is a result of the sharpness of the peak, so it is expected that for broader peaks the change will be less noticeable. In fact, we may plot the derivative of this phase with respect to the angle of incidence, as pictured in Figure \ref{fig,phase_d}(b), and notice that the maximum rates of changes are aligned with the peaks of transmitted power.

From a conceptual point of view, this rapidly changing phase results from the fact that a guided wave is excited in the plate, which is in itself a resonance phenomenon. This is yet another piece of evidence pointing to the relation between transmission peaks and guided wave excitation. Just as we expect a jump in phase by $\pi$ in multi-degree of freedom systems when it passes through a resonance, every time a peak in transmission is reached a new jump in phase will occur.

Of course this explanation is purely an attempt at making sense of the data present in Figure \ref{fig,phase_d}, so to get a more complete understanding of the behaviour of the phase, more work needs to be done on the subject. For now, since the phase does not seem to provide new information from the transmitted power, with regards to identifying peaks in transmission, and ultimately FTIR, we will end this discussion on this note. However, it needs to be stated that sharp peaks in transmission may be very hard to measure in practice without precise angular controls, meaning that, practically, these phase shifts can provide a complementary and more sensitive measure for their detection.

\section{Experimental Validation}
\label{sec,experimental_validation}
In this section we attempt to validate our theoretical model using experimental results. Instead of retaining the same mechanical system we have been using, here we consider an air-coupled setup as illustrated in Figure \ref{fig,validation}(a). A thin steel plate with a thickness of $d=\SI{50}{\micro\metre}$ is submerged in air on both sides. The steel plate has the properties of $c_{L2} = 5470$ m/s, $c_{S2} = 3093$ m/s and $\rho_2 = 7850$ kg/m\textsuperscript{3} and those of air are $c_{L1} = 332$ m/s and $\rho_1 = 1.3$ kg/m\textsuperscript{3} at the temperature of experiment. Two air-coupled transducers are placed on the two sides of the steel sample and are well adjusted to point to each other. One transducer generates an incident wave into air and the other receives the wave that travelled through the air-sample-air path between the two transducers. The angle $\theta$ between the incident wave and the sample surface is controlled by a stepper motor-driven rotary stage to which the sample is fixed.

\begin{figure}[!t]
    \centering
    \includegraphics[width=0.9\linewidth]{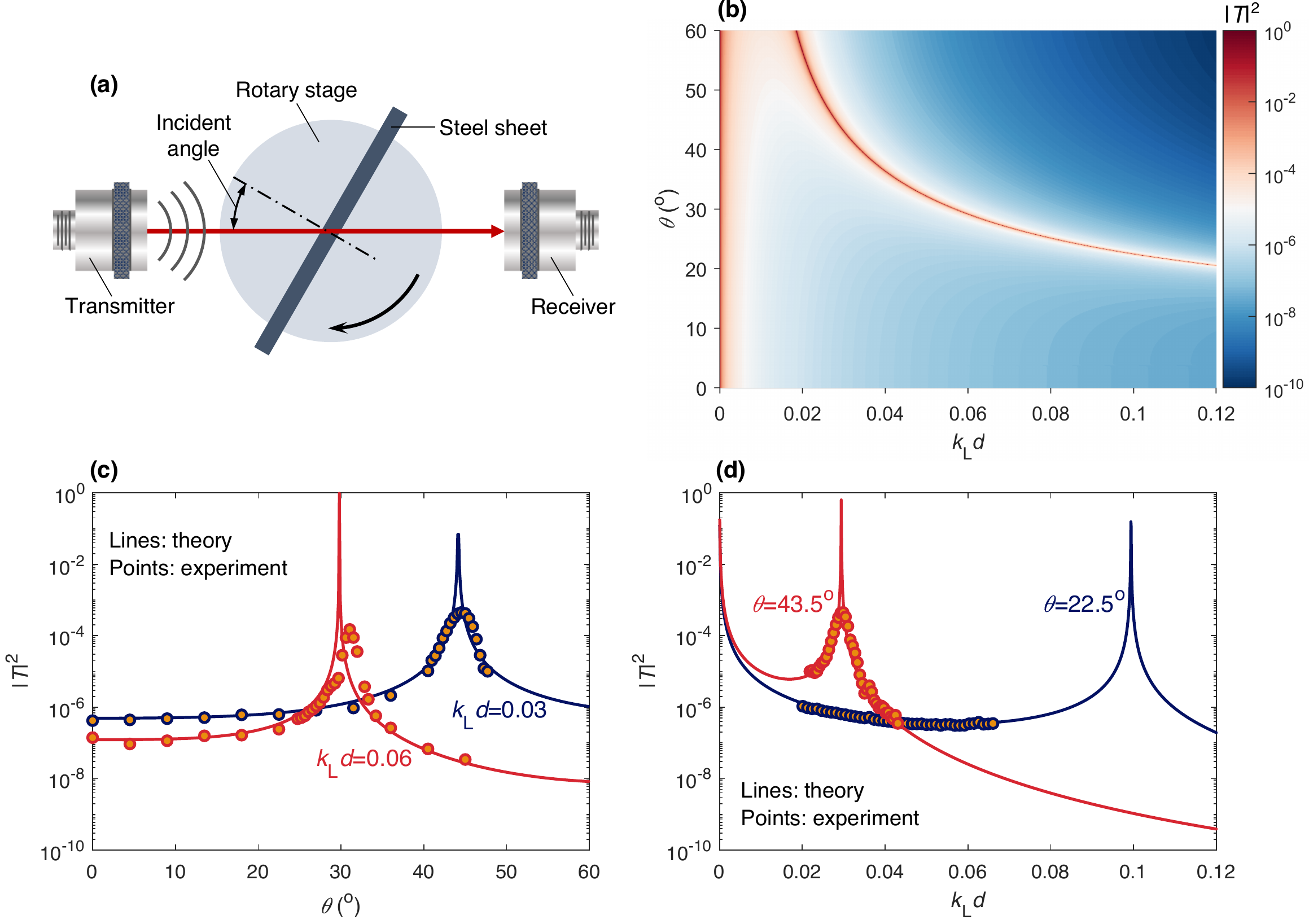}
    \caption{Experimental validation of the theory. (a) Experimental setup. A steel sample with a thickness of \SI{50}{\micro\metre} is placed in between two air-coupled transducers, with one for wave generation and another for receiving the wave transmitted through the sample. The angle between the incident wave and the sample surface is adjusted by the rotary stage on which the sample is fixed. (b) Theoretical result of transmitted power spectrum $|T|^2$ for the \SI{50}{\micro\metre} steel sample as a function of normalised frequency $k_{L}d$ and incident angle $\theta$. The normalised frequency is defined as the product between the thickness and the wavenumber of longitudinal waves inside the solid. (c) Comparison of theoretical and experimental results for the steel sample at the normalised frequencies of 0.03 and 0.06. (d) Similar comparison at the incident angles of 22.5$^\circ$ and 43.5$^\circ$.}
    \label{fig,validation}
\end{figure}

The theoretical power spectrum $|T|^2$ of the wave transmitted through the steel plate is given in Figure \ref{fig,validation}(b) as a function of normalised frequency $k_{L}d$ and incident angle $\theta$. Since the coupling between steel and air is very poor, it means that the resulting transmission coefficients are very low, and the guided wave modes are hard to visualise in a linear plot. To solve this issue, a logarithmic colour scale is used instead, allowing for a better definition of the guided wave modes in the transmitted power spectrum. The two critical angles for this system are $3.0^\circ$ and $6.4^\circ$. The fundamental antisymmetric mode $A_0$ of the plate is clearly highlighted in darker red colours, which occurs exactly in the FTIR regime beyond the two critical angles.

The experimental results are collected using three pairs of air-coupled transducers with the frequencies of 0.5, 0.7 and 1 MHz. For each transducer pair, a reference signal is acquired before placing the sample in between them, and then the transmitted signal is recorded at each angle of incidence when the sample is present. For each incident angle, the resulting transmitted signal is divided by the reference signal in the frequency domain to obtain the transmission coefficient over a frequency range. Combining the three transducer pairs with different incident angles, the experimental results are obtained to form a transmitted power map similar to that of Figure \ref{fig,validation}(b). But for brevity, the experiment-theory comparison is only provided for two sets of representative results in Figure \ref{fig,validation}(c) and (d), showing respectively the angle dependence of the transmitted power at $k_{L}d=0.03$ and 0.06 and the frequency dependence at $\theta=22.5^\circ$ and $43.5^\circ$.

It can be seen from Figure \ref{fig,validation}(c) and (d) that the theoretical predictions and the experimental values match very well, even in such an extreme case of barely any coupling, between steel and air. The theory accurately predicts the fine details at, and around, the peaks that are exposed in the experimental results. To further complement the validation, we have also carried out the same experiments on two other samples of 250 and \SI{500}{\micro\metre} thickness, and the results substantiated the good agreement.

\section{Conclusion}
\label{sec,conclusion}
In this work we hoped to frame FTIR in a theoretical basis, as a possible effect to be harnessed when using applications reliant on the propagation of elastic waves. The success of the potentials-based model is apparent when compared to the experimental data gathered, as it provides a method to adequately predict in what conditions FTIR can occur.

Besides that, the objective of understand the physical mechanisms responsible for FTIR was also achieved, as its connection with the existence of guided waves in the plate emerges naturally from the theoretical formulation. In fact, one could conclude that FTIR is caused by two main factors, based on the transmitted power plot shown in Figure \ref{fig,dispersion_curves}.

First, FTIR can occur for very thin plates ($d \ll \lambda_{L1}$), just like in the case of optics. This is a direct result of the exponential decay of the evanescent wave inside the plate. If the thickness is sufficiently small, this decay will not be enough to decrease the amplitude of the evanescent wave when it reaches the other plate boundary to a negligible value, which means that a non zero wave is going to be transmitted into the fluid on the other side of the plate.

However, and this is perhaps the most interesting conclusion of this work, FTIR can also happen when the incident wave is such that it excites the fundamental antisymmetric mode $A_0$ of the plate. This other way of enabling FTIR is not so restrictive in terms of the plate thickness (even happening for $d > 2\lambda_{L1}$ in the water-copper-water system), although it requires a specific angle of incidence for the peak to appear. We could even go further, based on the work of Chimenti and Rokhlin \cite{Chimenti and Rokhlin_1990}, and state that for rarer fluids this mode would extend for even thicker plates, which, as we approach the limit of a plate surrounded by a vacuum, would make FTIR a global phenomenon.

One could also argue that it may be possible to have FTIR when the fundamental symmetric mode $S_0$ is excited. This is true in part, as for thicker plates the guided wave velocity of the fundamental extensional mode approaches that of the fundamental flexural mode of the plate. It is possible to observe this region in Figure \ref{fig,dispersion_curves}, however due to its limited applicability, it is safer to consider $A_0$ as the main cause of FTIR for thicker plates.

\section{Acknowledgements}
\label{sec,acknowledgements}
This work is supported by the EPSRC CDT in Future Innovation in NDE. Bo Lan gratefully acknowledges the Imperial College Research Fellowship and Ming Huang the generous funding from the NDE group at Imperial College London.

\appendix
\section{Multiple reflections approach using ray-based model}
\label{append,multiple_reflections}
To establish a multiple reflections approach, we will follow a formulation analogous to the one used by Zhu et al \cite{Zhu et al_1986} for light. Figure \ref{fig,multiple_reflections} is a representation of all the possible pathways a ray can travel, if a single reflection inside the plate is considered. Throughout this discussion, media 1 and 3 are considered fluids and medium 2 is an elastic isotropic solid plate.

\begin{figure}[!t]
    \centering
    \includegraphics[scale=0.6]{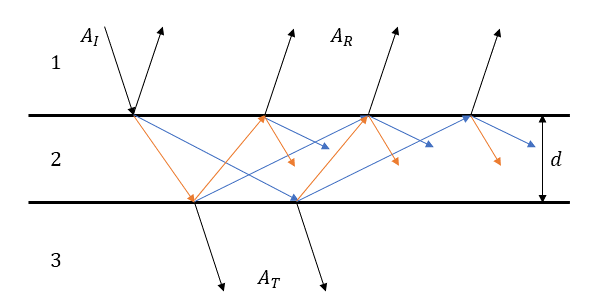}
    \caption{Representation of the multiply reflected wave pathways inside the plate. Blue arrows represent longitudinal, while orange ones represent shear waves. The first and third region represent a liquid, while region 2 is a solid plate of thickness $d$.}
    \label{fig,multiple_reflections}
\end{figure}

In this figure, one of the main complexities that needs to be accounted for when working with elastic waves is easily seen. Every time a wave interacts with a boundary, both reflection and transmission occur, and one longitudinal and one shear waves are produced for each, unless the wave escapes into the liquid. This fact alone makes this problem severely more complex than the case of electromagnetic waves, where the two different polarisations are completely uncoupled.

The ratio of reflected and transmitted energies are controlled by two factors. First, when a wave interacts with a boundary, other waves will be produced, so equations equivalent to the Fresnel relations \cite{Born and Wolf_1980} need to be defined for all possible interactions. There is of course, as mentioned before, extensive literature on the subject, such as \cite{Graff_1975,Rose_1999}, so this section will focus on establishing the sums that will result in a formula for the total transmitted and reflected waves.

The second factor is interfering waves inside the plate, due to phase differences between them, that will result in different amounts of reflected and transmitted energies. To account for this, as mentioned in the main text, we need to calculate the contribution of each single ray path to the total reflected or transmitted wave.

For the sake of simplicity, let us assume that both fluids are equal. Also, by conservation of energy, since the plate will eventually radiate all those trapped waves, the sum between the reflected and transmitted power should be equal to 1. This means that it is possible to calculate only the sum of contributions to the total reflected wave, and still have information about the transmitted power by subtracting the reflected power from 1.

Then, it is imperative to define the following boundary interaction factors: $r_{12}$ reflection of waves inside the fluid, $l_{12}$ and $s_{12}$ transmission into a longitudinal or shear wave inside the solid, respectively, $ll_{21}$ and $ls_{21}$ transmission into the liquid from an incident longitudinal or shear wave in the solid, respectively, $rl_{21}$ and $rs_{21}$ reflection of longitudinal or shear waves inside the solid, respectively, and finally $ml_{21}$ and $ms_{21}$ mode conversion between a longitudinal and shear wave, and its reverse, in the solid, respectively. Besides that, we need also to define the path difference or phase shift that each particular wave will be subjected to \cite{Zhu et al_1986},

\begin{equation}
\label{eqn,phase_shift_longitudinal}
\delta_L = 2\pi fd\sqrt{\frac{1}{c_{L2}^2}-\frac{1}{c_{L1}^2}\sin^2\theta}
\end{equation}

\begin{equation}
\label{eqn,phase_shift_shear}
\delta_S = 2\pi fd\sqrt{\frac{1}{c_{S2}^2}-\frac{1}{c_{L1}^2}\sin^2\theta}
\end{equation}

\noindent where $\delta_L$ and $\delta_S$ are the phase changes longitudinal and shear waves undergo between the two ends of the plate, respectively, and $\theta$ is the angle of incidence of $A_I$. All the other variables were already defined in the main text.

To understand the sequence of contributions we calculated all the possible paths for a number of multiple reflections and tried to figure out the nth term of the resulting mathematical series. Since such an extensively detailed work would be very cumbersome to write fully here, we will discuss a single case before showing the complete series for the sum of all possible paths inside the plate. We invite readers with interest in this to try other cases and check, as we did privately, the validity of our general equation.

Before starting, we need to state three more quantities that will be useful in terms of defining each path and the number of terms in the infinite sum. $N$ will be half the total number of paths considered, which ideally will tend to infinity. As an example, Figure \ref{fig,multiple_reflections} is a representation of $N=1$, since each reflection contains two paths. $n$ will be the number of shear paths a wave takes through a certain pathway. This means that a longitudinal wave reflecting as itself will always have $n=0$, while the inverse can be said for shear waves $n=2N$. Finally, $k$ will be the number of mode conversions that happen in a certain pathway. Having defined these, let us study all the possible paths for $N=5$, where each type of path is defined by $(n,k)$,

\begin{align*}
&A_R = l_{12}ll_{21}rl_{21}^9e^{10i\delta_L} &:(0,0)\\
&A_R = l_{12}ls_{21}ml_{21}rl_{21}^8e^{9i\delta_L}e^{i\delta_S} + s_{12}ll_{21}ms_{21}rl_{21}^8e^{9i\delta_L}e^{i\delta_S} &:(1,1)\\
&A_R = 8l_{12}ll_{21}ml_{21}ms_{21}rl_{21}^7e^{9i\delta_L}e^{i\delta_S} &:(1,2)\\
&A_R = l_{12}ls_{21}ml_{21}rl_{21}^7rs_{21}e^{8i\delta_L}e^{2i\delta_S} + s_{12}ll_{21}ms_{21}rl_{21}^7rs_{21}e^{8i\delta_L}e^{2i\delta_S} &:(2,1)\\
&A_R = 7l_{12}ll_{21}ml_{21}ms_{21}rl_{21}^6rs_{21}e^{8i\delta_L}e^{2i\delta_S} + s_{12}ls_{21}ml_{21}ms_{21}rl_{21}^7e^{8i\delta_L}e^{2i\delta_S} &:(2,2)\\
&A_R = 7l_{12}ls_{21}ml_{21}^2ms_{21}rl_{21}^6e^{8i\delta_L}e^{2i\delta_S} + 7s_{12}ll_{21}ml_{21}ms_{21}^2rl_{21}^6e^{8i\delta_L}e^{2i\delta_S} &:(2,3)\\
&A_R = 21l_{12}ll_{21}ml_{21}^2ms_{21}^2rl_{21}^5e^{8i\delta_L}e^{2i\delta_S} &:(2,4)\\
&A_R = l_{12}ls_{21}ml_{21}rl_{21}^6rs_{21}^2e^{7i\delta_L}e^{3i\delta_S} + s_{12}ll_{21}ms_{21}rl_{21}^6rs_{21}^2e^{7i\delta_L}e^{3i\delta_S} &:(3,1)\\
&A_R = 6l_{12}ll_{21}ml_{21}ms_{21}rl_{21}^5rs_{21}^2e^{7i\delta_L}e^{3i\delta_S} + 2s_{12}ls_{21}ml_{21}ms_{21}rl_{21}^6rs_{21}e^{7i\delta_L}e^{3i\delta_S} &:(3,2)\\
&A_R = 12l_{12}ls_{21}ml_{21}^2ms_{21}rl_{21}^5rs_{21}e^{7i\delta_L}e^{3i\delta_S} + 12s_{12}ll_{21}ml_{21}ms_{21}^2rl_{21}^5rs_{21}e^{7i\delta_L}e^{3i\delta_S} &:(3,3)\\
&A_R = 30l_{12}ll_{21}ml_{21}^2ms_{21}^2rl_{21}^4rs_{21}e^{7i\delta_L}e^{3i\delta_S} + 6s_{12}ls_{21}ml_{21}^2ms_{21}^2rl_{21}^5e^{7i\delta_L}e^{3i\delta_S} &:(3,4)\\
&A_R = 15l_{12}ls_{21}ml_{21}^3ms_{21}^2rl_{21}^4e^{7i\delta_L}e^{3i\delta_S} + 15s_{12}ll_{21}ml_{21}^2ms_{21}^3rl_{21}^4e^{7i\delta_L}e^{3i\delta_S} &:(3,5)\\
&A_R = 20l_{12}ll_{21}ml_{21}^3ms_{21}^3rl_{21}^3e^{7i\delta_L}e^{3i\delta_S} &:(3,6)\\
&A_R = l_{12}ls_{21}ml_{21}rl_{21}^5rs_{21}^3e^{6i\delta_L}e^{4i\delta_S} + s_{12}ll_{21}ms_{21}rl_{21}^5rs_{21}^3e^{6i\delta_L}e^{4i\delta_S} &:(4,1)\\
&A_R = 5l_{12}ll_{21}ml_{21}ms_{21}rl_{21}^4rs_{21}^3e^{6i\delta_L}e^{4i\delta_S} + 3s_{12}ls_{21}ml_{21}ms_{21}rl_{21}^5rs_{21}^2e^{6i\delta_L}e^{4i\delta_S} &:(4,2)\\
&A_R = 15l_{12}ls_{21}ml_{21}^2ms_{21}rl_{21}^4rs_{21}^2e^{6i\delta_L}e^{4i\delta_S} + 15s_{12}ll_{21}ml_{21}ms_{21}^2rl_{21}^4rs_{21}^2e^{6i\delta_L}e^{4i\delta_S} &:(4,3)\\
&A_R = 30l_{12}ll_{21}ml_{21}^2ms_{21}^2rl_{21}^3rs_{21}^2e^{6i\delta_L}e^{4i\delta_S} + 15s_{12}ls_{21}ml_{21}^2ms_{21}^2rl_{21}^4rs_{21}e^{6i\delta_L}e^{4i\delta_S} &:(4,4)\\
&A_R = 30l_{12}ls_{21}ml_{21}^3ms_{21}^2rl_{21}^3rs_{21}e^{6i\delta_L}e^{4i\delta_S} + 30s_{12}ll_{21}ml_{21}^2ms_{21}^3rl_{21}^3rs_{21}e^{6i\delta_L}e^{4i\delta_S} &:(4,5)\\
&A_R = 30l_{12}ll_{21}ml_{21}^3ms_{21}^3rl_{21}^2rs_{21}e^{6i\delta_L}e^{4i\delta_S} + 10s_{12}ls_{21}ml_{21}^3ms_{21}^3rl_{21}^3e^{6i\delta_L}e^{4i\delta_S} &:(4,6)\\
&A_R = 10l_{12}ls_{21}ml_{21}^4ms_{21}^3rl_{21}^2e^{6i\delta_L}e^{4i\delta_S} + 10s_{12}ll_{21}ml_{21}^3ms_{21}^4rl_{21}^2rs_{21}e^{6i\delta_L}e^{4i\delta_S} &:(4,7)\\
&A_R = 5l_{12}ll_{21}ml_{21}^4ms_{21}^4rl_{21}e^{6i\delta_L}e^{4i\delta_S} &:(4,8)\\
&A_R = l_{12}ls_{21}ml_{21}rl_{21}^4rs_{21}^4e^{5i\delta_L}e^{5i\delta_S} + s_{12}ll_{21}ms_{21}rl_{21}^4rs_{21}^4e^{5i\delta_L}e^{5i\delta_S} &:(5,1)\\
&A_R = 4l_{12}ll_{21}ml_{21}ms_{21}rl_{21}^3rs_{21}^4e^{5i\delta_L}e^{5i\delta_S} + 4s_{12}ls_{21}ml_{21}ms_{21}rl_{21}^4rs_{21}^3e^{5i\delta_L}e^{5i\delta_S} &:(5,2)\\
&A_R = 16l_{12}ls_{21}ml_{21}^2ms_{21}rl_{21}^3rs_{21}^3e^{5i\delta_L}e^{5i\delta_S} + 16s_{12}ll_{21}ml_{21}ms_{21}^2rl_{21}^3rs_{21}^3e^{5i\delta_L}e^{5i\delta_S} &:(5,3)\\
&A_R = 24l_{12}ll_{21}ml_{21}^2ms_{21}^2rl_{21}^2rs_{21}^3e^{5i\delta_L}e^{5i\delta_S} + 24s_{12}ls_{21}ml_{21}^2ms_{21}^2rl_{21}^3rs_{21}^2e^{5i\delta_L}e^{5i\delta_S} &:(5,4)\\
&A_R = 36l_{12}ls_{21}ml_{21}^3ms_{21}^2rl_{21}^2rs_{21}^2e^{5i\delta_L}e^{5i\delta_S} + 36s_{12}ll_{21}ml_{21}^2ms_{21}^3rl_{21}^2rs_{21}^2e^{5i\delta_L}e^{5i\delta_S} &:(5,5)\\
&A_R = 24l_{12}ll_{21}ml_{21}^3ms_{21}^3rl_{21}rs_{21}^2e^{5i\delta_L}e^{5i\delta_S} + 24s_{12}ls_{21}ml_{21}^3ms_{21}^3rl_{21}^2rs_{21}e^{5i\delta_L}e^{5i\delta_S} &:(5,6)\\
&A_R = 16l_{12}ls_{21}ml_{21}^4ms_{21}^3rl_{21}rs_{21}e^{5i\delta_L}e^{5i\delta_S} + 16s_{12}ll_{21}ml_{21}^3ms_{21}^4rl_{21}rs_{21}e^{5i\delta_L}e^{5i\delta_S} &:(5,7)\\
&A_R = 4l_{12}ll_{21}ml_{21}^4ms_{21}^4rs_{21}e^{5i\delta_L}e^{5i\delta_S} + 4s_{12}ls_{21}ml_{21}^4ms_{21}^4rl_{21}e^{5i\delta_L}e^{5i\delta_S} &:(5,8)\\
&A_R = l_{12}ls_{21}ml_{21}^5ms_{21}^4e^{5i\delta_L}e^{5i\delta_S} + s_{12}ll_{21}ml_{21}^4ms_{21}^5e^{5i\delta_L}e^{5i\delta_S} &:(5,9)
\end{align*}

\noindent where the same trend continues from $n=6$ until $n=10$, exchanging longitudinal paths with shear paths.

\begin{figure}[!t]
    \centering
    \includegraphics[scale=0.6]{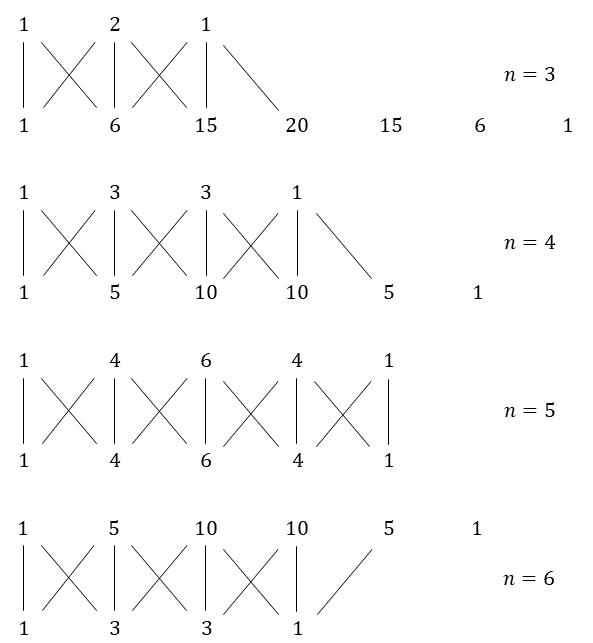}
    \caption{Schematic of the sequence of path coefficients for a specific number of total paths, $N=5$, using rows of Pascal's triangle. Lines represent multiplications between two binomial coefficients. The vertical lines give the coefficients of the Odd Paths, while diagonal lines which cross each other give the Even Paths. The isolated diagonal lines at the edge of the diagrams correspond to the coefficients of the End Paths. All the diagrams from $n=0$ to $n=10$ need to be considered.}
    \label{fig,pascal_triangle}
\end{figure}

Naturally, the coefficients shown seem a bit random at first, but using Pascal's triangle, it is clear from Figure \ref{fig,pascal_triangle} that they are a result of multiplication between binomial coefficients. Based on the information in Figure \ref{fig,pascal_triangle}, we can subdivide the paths into three different categories, since it is not possible to find a general term for every available path. The vertical lines in Figure \ref{fig,pascal_triangle} represent what we call the Odd Paths. The crosses represent the Even Paths. Finally, the single diagonal lines represent the End Paths. Grouping the paths into these three groups allows for a series representation of the infinite sum of contributions for calculating the total reflected wave.

Therefore, for the End, Odd, and Even Paths we have, respectively,

\begin{multline}
\label{eqn,end_paths}
\left(\frac{A_R}{A_I}\right)_{End} = r_{12} + \sum_{N=1}^{\infty}\sum_{n=0}^{N-1}\binom{2N-n-1}{n}\left[l_{12}ll_{21}ml_{21}^nms_{21}^nrl_{21}^{2N-2n-1}e^{i(2N-n)\delta_L}e^{in\delta_S} + \right.\\\left. s_{12}ls_{21}ml_{21}^nms_{21}^nrs_{21}^{2N-2n-1}e^{in\delta_L}e^{i(2N-n)\delta_S}\right]
\end{multline}

\begin{multline}
\label{eqn,odd_paths}
\left(\frac{A_R}{A_I}\right)_{Odd} = \sum_{N=1}^{\infty}\sum_{n=1}^{2N-1}\sum_{k=1}^{N-|n-N|}\binom{2N-n-1}{k-1}\binom{n-1}{k-1}\left[l_{12}ls_{21}ml_{21}^kms_{21}^{k-1}rl_{21}^{2N-n-k}rs_{21}^{n-k} + \right.\\\left. s_{12}ll_{21}ml_{21}^{k-1}ms_{21}^krl_{21}^{2N-n-k}rs_{21}^{n-k}\right]e^{i(2N-n)\delta_L}e^{in\delta_S}
\end{multline}

\begin{multline}
\label{eqn,even_paths}
\left(\frac{A_R}{A_I}\right)_{Even} = \sum_{N=2}^{\infty}\sum_{n=2}^{2N-2}\sum_{k=1}^{N-|n-N|-1}\left[\binom{2N-n-1}{k}\binom{n-1}{k-1}l_{12}ll_{21}ml_{21}^kms_{21}^krl_{21}^{2N-n-k-1}rs_{21}^{n-k} + \right.\\\left. \binom{2N-n-1}{k-1}\binom{n-1}{k}s_{12}ls_{21}ml_{21}^kms_{21}^krl_{21}^{2N-n-k}rs_{21}^{n-k-1}\right]e^{i(2N-n)\delta_L}e^{in\delta_S}
\end{multline}

\noindent and to obtain the total reflected wave $A_R$ we just need to sum the results of these three equations.

Equations (\ref{eqn,end_paths}) to (\ref{eqn,even_paths}) are already enough for a finite sum representation of the system. The results shown in Figure \ref{fig,partial_d} use such a finite number of terms, with $N=40$. However, the sum is not very well-behaved for a large number of paths, which makes an analytical solution much more desirable.

Looking at the fact that we have an infinite sum of powers weighted by binomial coefficients, a possible solution should be given in terms of hypergeometrical functions \cite{Hayden and Lamagna_1986}. We managed to find a closed-form representation of the first sum, in equation (\ref{eqn,end_paths}). For the other two, there is still no other more simplified way of representing them, which is one of the reasons why we ended up dropping this formulation in favor of the potentials-based one. However, for the sake of completeness, let us develop a closed-form solution for the sum of the contributions of the End Paths.

Let us start by writing the sum in equation (\ref{eqn,end_paths}) in a more suggestive form,

\begin{multline}
\label{eqn,closed_form_derivation_1}
\left(\frac{A_R}{A_I}\right)_{End} = r_{12} + l_{12}ll_{21}e^{i\delta_L}\sum_{N=1}^{\infty}\left[\left(rl_{21}e^{i\delta_L}\right)^{2N-1}\sum_{n=0}^{N-1}\binom{2N-n-1}{n}\left(\frac{ml_{21}ms_{21}e^{i\delta_S}}{rl_{21}^2e^{i\delta_L}}\right)^n\right] + \\
s_{12}ls_{21}e^{i\delta_S}\sum_{N=1}^{\infty}\left[\left(rs_{21}e^{i\delta_S}\right)^{2N-1}\sum_{n=0}^{N-1}\binom{2N-n-1}{n}\left(\frac{ml_{21}ms_{21}e^{i\delta_L}}{rs_{21}^2e^{i\delta_S}}\right)^n\right]
\end{multline}

According to Hayden and Lamagna \cite{Hayden and Lamagna_1986}, the following sum can be represented as,

\begin{equation}
\label{eqn,closed_form_derivation_2}
\sum_{n=0}^{N-1}\binom{2N-n-1}{n}x^n = \prescript{}{2}{F}_1\left(\frac{1}{2}-N,1-N,1-2N,-4x\right)
\end{equation}

\noindent where $\prescript{}{2}{F}_1$ is the hypergeometric function defined as \cite{Abramowitz and Stegun_1972},

\begin{equation}
\label{eqn,hypergeometric_function}
\prescript{}{2}{F}_1\left(a,b,c,z\right) = \sum_{n=0}^\infty\frac{(a)_n(b)_n}{(c)_n}\frac{z^n}{n!}
\end{equation}

\noindent and $(\cdot)_n$ is the rising Pochhammer symbol also called rising factorial \cite{Abramowitz and Stegun_1972},

\begin{equation}
\label{eqn,rising_factorial}
(a)_n = \begin{cases}
1\quad\quad\quad\quad\quad\quad\quad\quad\quad\,,\,n=0 \\
a(a+1) ... (a+n-1)\,\,,\,n>0
\end{cases}
\end{equation}

Besides this, according to Abramowitz and Stegun \cite{Abramowitz and Stegun_1972}, the following hypergeometric function has the closed-form representation,

\begin{equation}
\label{eqn,closed_form_derivation_3}
\prescript{}{2}{F}_1\left(\frac{1}{2}-N,1-N,1-2N,-4x\right) = \frac{1}{\sqrt{1+4x}}\left[\frac{1}{2}\left(1+\sqrt{1+4x}\right)\right]^{2N}
\end{equation}

Since both the sums of equation (\ref{eqn,closed_form_derivation_1}) have exactly the form of equation (\ref{eqn,closed_form_derivation_2}), we can write directly their results as in equation (\ref{eqn,closed_form_derivation_3}),

\begin{multline}
\label{eqn,closed_form_derivation_4}
\left(\frac{A_R}{A_I}\right)_{End} = r_{12} + \frac{l_{12}ll_{21}}{rl_{21}\sqrt{1+4\frac{ml_{21}ms_{21}e^{i\delta_S}}{rl_{21}^2e^{i\delta_L}}}} \frac{\left(1+\sqrt{1+4\frac{ml_{21}ms_{21}e^{i\delta_S}}{rl_{21}^2e^{i\delta_L}}}\right)^2rl_{21}^2e^{2i\delta_L}}{4-\left(1+\sqrt{1+4\frac{ml_{21}ms_{21}e^{i\delta_S}}{rl_{21}^2e^{i\delta_L}}}\right)^2rl_{21}^2e^{2i\delta_L}} + \\ \frac{s_{12}ls_{21}}{rs_{21}\sqrt{1+4\frac{ml_{21}ms_{21}e^{i\delta_L}}{rs_{21}^2e^{i\delta_S}}}} \frac{\left(1+\sqrt{1+4\frac{ml_{21}ms_{21}e^{i\delta_L}}{rs_{21}^2e^{i\delta_S}}}\right)^2rs_{21}^2e^{2i\delta_S}}{4-\left(1+\sqrt{1+4\frac{ml_{21}ms_{21}e^{i\delta_L}}{rs_{21}^2e^{i\delta_S}}}\right)^2rs_{21}^2e^{2i\delta_S}}
\end{multline}

\noindent which is the final closed-form representation for the sum of the End Paths.

This means that, with the simplification of equation (\ref{eqn,closed_form_derivation_4}), the full reflected wave takes the form,

\begin{equation}
\label{eqn,closed_form_derivation_5}
\frac{A_R}{A_I} = r_{12} + \frac{l_{12}ll_{21}}{\sqrt{1+M_l}} \frac{\left(1+\sqrt{1+M_l}\right)^2rl_{21}e^{2i\delta_L}}{4-\left(1+\sqrt{1+M_l}\right)^2rl_{21}^2e^{2i\delta_L}} + \frac{s_{12}ls_{21}}{\sqrt{1+M_s}}
\frac{\left(1+\sqrt{1+M_s}\right)^2rs_{21}e^{2i\delta_S}}{4-\left(1+\sqrt{1+M_s}\right)^2rs_{21}^2e^{2i\delta_S}} + M
\end{equation}

\noindent with,

\begin{equation}
\label{eqn,closed_form_derivation_6}
M_l = 4\frac{ml_{21}ms_{21}e^{i\delta_S}}{rl_{21}^2e^{i\delta_L}},\; M_s = 4\frac{ml_{21}ms_{21}e^{i\delta_L}}{rs_{21}^2e^{i\delta_S}}, \; M = \left(\frac{A_R}{A_I}\right)_{Odd} + \left(\frac{A_R}{A_I}\right)_{Even}
\end{equation}

Finally, if we consider that no mode conversion occurs ($ml_{21}=ms_{21}=0$, or else $M_l=M_s=M=0$), which is to say that we are only interested in reverberations of longitudinal and shear waves, we get the following equation,

\begin{equation}
\label{eqn,closed_form_derivation_7}
\frac{A_R}{A_I} = r_{12} + l_{12}ll_{21}\frac{rl_{21}e^{2i\delta_L}}{1-rl_{21}^2e^{2i\delta_L}} + s_{12}ls_{21}\frac{rs_{21}e^{2i\delta_S}}{1-rs_{21}^2e^{2i\delta_S}} 
\end{equation}

\noindent which is a much simpler equation and comparable to the results obtained by Zhu et al \cite{Zhu et al_1986}.

In fact, equation (\ref{eqn,closed_form_derivation_7}) would be the geometric sum obtained if longitudinal and shear waves did not interact with one another, which is what happens between light polarisations. Such a result is encouraging, and confirms that the fundamental assumptions used in this multiple reflections approach are indeed correct. However, for practical reasons, and since we were interested in studying the behaviour of evanescent waves, ultimately this methodology was deemed insufficient.

\end{document}